\documentclass{ws-ijmpa}

\usepackage[super]{cite}
\usepackage{xcolor}
\usepackage[verbose,hypertexnames=false]{hyperref}
\hypersetup{colorlinks=false,allbordercolors=blue,pdfborderstyle={/S/U/W 1}}

\begin{document}

\markboth{Zhiqing Zhang}{Overview of tau lepton physics at a super tau-charm facility}

%
\catchline{}{}{}{}{}
%

\title{Overview of tau lepton physics at super tau-charm factory\footnote{Invited plenary talk.}
}

\author{Zhiqing Zhang
}

\address{Univ.\ Paris-Saclay, IJCLab, IN2P3/CNRS, 91405 Orsay Cedex, France
\\
Zhiqing.Zhang@ijclab.in2p3.fr}



\maketitle

\begin{history}
\received{Day Month Year}
\revised{Day Month Year}
\accepted{Day Month Year}
\published{Day Month Year}
\end{history}

\begin{abstract}
An overview of tau lepton physics is presented, using the tau lepton discovery and its precision measurements as examples to illustrate the importance of the energy region to be covered by the super tau-charm factory. By presenting the current measurement status of the major physics topics, the emphasize is put on pointing out a few open issues and possibilities for the super tau-charm factory.
\end{abstract}


\ccode{PACS numbers: 13.35.Dx, 14.60.Fg}

\section{Introduction}	
The year 2025 is a special year as it marks the 50th anniversary of the tau lepton discovery~\cite{PhysRevLett.35.1489} and the 25th anniversary of the 1st direct evidence for the tau neutrino~\cite{DONUT:2000fbd}.
Tau lepton physics at the super tau-charm facility/factory (STCF) has been discussed in several excellent talks in the previous workshops, see e.g.\ Ref.~\cite{Pich:2024qob}.
Instead of repeating this, I would rather give a few examples from the discovery to precision measurements and comment on these and point out a few open issues or possibilities. Whenever possible, I would use materials that I have personally obtained.

The tau lepton and tau neutrino are the third generation leptons in the standard model (SM). The SM describing the elementary particles and their interactions has been extremely successful. Yet, it is believed that there must be physics beyond the SM (BSM). 
The question is where to find new physics. A direct discovery of new physics may generally need a higher energy collider. The tau lepton discovery itself is a good example of this. The high precision frontier is the other possibility. The STCF belongs to the latter category. 

\section{Tau lepton discovery and its properties}
\subsection{The discovery}
The tau lepton was discovered in 1975 by the Mark\,I experiment at the SPEAR $e^+e^-$ storage ring. The collider started its operation in 1973 and was the highest energy collider back then. The Mark\,I detector was one of the first large solid angle, general purpose detectors built for colliding beams. The second-generation charm quark was discovered in 1974 with the same detector and collider. 

The search for a sequential heavy lepton had the theoretical guidance of Refs.~\cite{PhysRevD.4.2821,PhysRevD.7.887}.
The discovery was based initially on 24 events of the type $e^+e^-\to e^\pm\mu^\mp +\text{missing energy}$, that could not be explained by hadron misidentification into leptons and any other known processes. The new particle was initially called $U$ for unknown and was named later as $\tau$ from Greek for third. The discovery was confirmed in 1977 by the PLUTO experiment at DORIS~\cite{BURMESTER1977297}.  

\begin{figure}[hbt]
\centerline{\includegraphics[width=4.5cm]{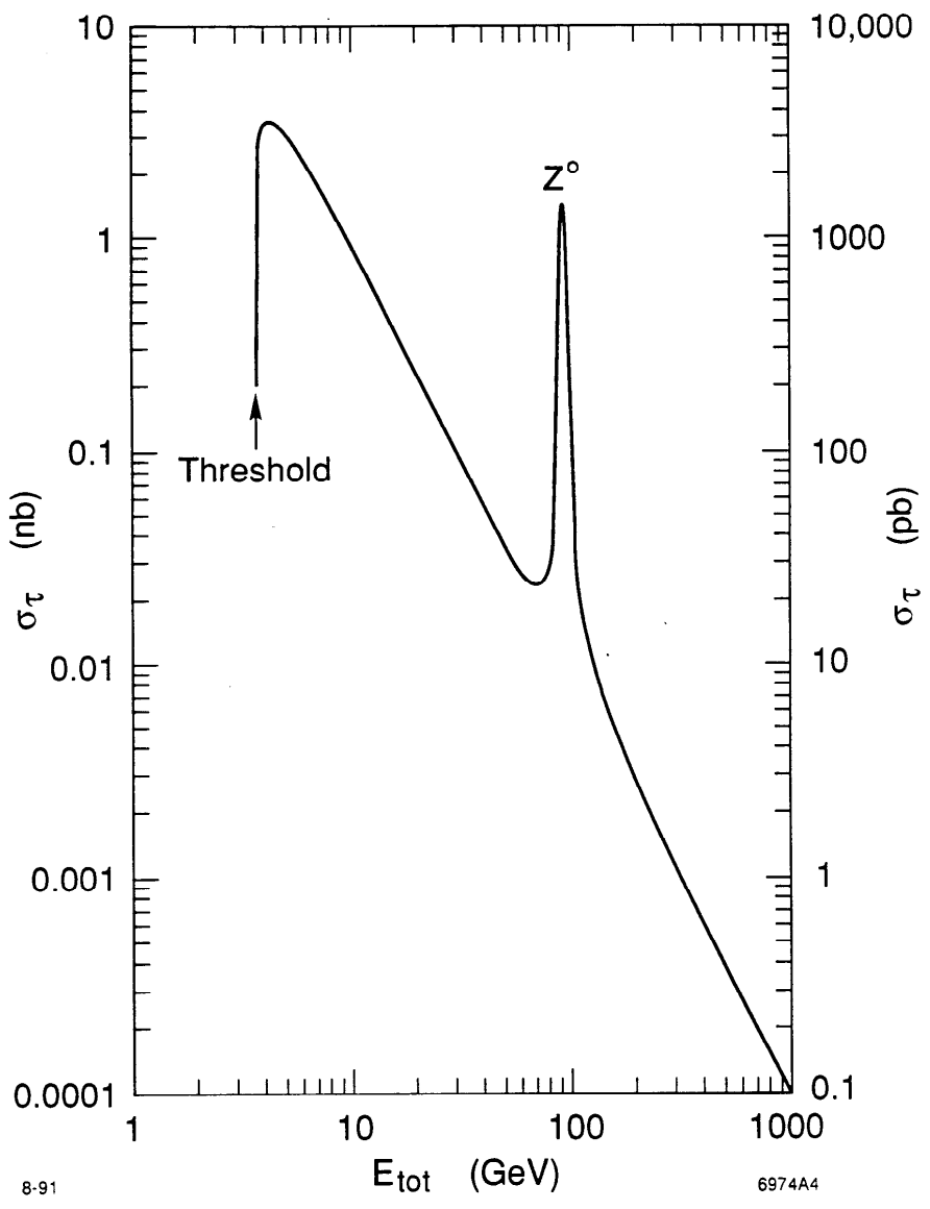}}
\caption{Tau pair production cross section as a function of center-of-mass energy. Figure taken from Ref.~\cite{Perl:1991gd}.}
\label{fig:xsec}
\end{figure}

It turned out that the SPEAR collider was running at the right energy since the tau production cross section is about the largest at around 4.5\,GeV (Figure~\ref{fig:xsec}). This illustrates that a discovery needs not only the sufficient (high) energy but also to be in the right energy range.

The energy range proposed by the STCF will cover both the threshold and highest cross section regions of tau lepton pair production. 
In comparison, the production cross sections at a $B$-factory around 10.58\,GeV and a $Z$ factory will be lower. 
The event rate depends, however, not only the production cross section but also the luminosity of a collider.
SuperKEKB has reached a world record of $0.5\times 10^{35}\,\mathrm{cm}^{-2}\mathrm{s}^{-1}$ in 2024 and is aiming for an improvement for another order of magnitude to achieve an integrated luminosity of 50\,ab$^{-1}$. In this sense, the SuperKEKB is an existing super tau factory.

\subsection{Precision property measurements}
At the future STCF, one can perform precision measurements. The tau mass measurement is a good example. Precision measurements are important since they are needed to improve the precision of fundamental SM parameters that are currently limiting factors for testing the SM and constraining new physics. Precision measurements are also needed for finding deviations from the SM predictions, for performing lepton-flavor-universality (LFU) tests, for searching for 
 lepton-flavor-violating (LFV), for measuring tau electric dipole moment ($d_\tau$) and tau magnetic moment anomaly ($a_\tau$), and for investigating CP violation in tau decays.

\subsubsection{Tau lepton mass measurements}
Before 1992, the mass of the tau lepton ($m_\tau$) was known with an uncertainty of 3--4\,MeV/c$^2$. 
The first measurement from the BES experiment using 14 $e\mu$ events from tau pair production near threshold has greatly improved its precision to well below 1\,MeV/c$^2$: $1776.9^{+0.4}_{-0.5}\pm 0.2$\,MeV/c$^2$~\cite{PhysRevLett.69.3021}, where the uncertainties are statistical and systematic, respectively. The measurement was improved in 1996 to $1776.96^{+0.18}_{-0.21}$$^{+0.25}_{-0.17}$\,MeV/c$^2$ by including several other decay channels~\cite{BES:1995jmv} and later in 2014 to $1776.91\pm 0.12^{+0.10}_{-0.13}$\,MeV/c$^2$ based on about 1200 tau pair events in 13 decay channels over 4 energy scan points (one below and three above the production threshold) selected from a data sample recorded with the new BESIII detector, corresponding to an integrated luminosity of about 24\,pb$^{-1}$~\cite{BESIII:2014srs}. The systematic uncertainty of the latter is dominated by the fitted efficiency uncertainty of $^{+0.038}_{-0.034}$\,MeV/c$^2$, background shape uncertainty of $\pm 0.040$\,MeV/c$^2$, selection and misidentification uncertainties of $\pm 0.050$ and $\pm 0.048$\,MeV/c$^2$, beam energy scale uncertainty of $^{+0.022}_{-0.086}$\,MeV/c$^2$, and beam energy spread uncertainty of $\pm 0.016$\,MeV/c$^2$. These measurements demonstrated the advantage of the (S)TCF in measuring $m_\tau$.

It is interesting to compare with the most recent measurement from Belle\,II at SuperKEKEB, giving $1777.09\pm 0.08\pm 0.11$\,MeV/c$^2$~\cite{Belle-II:2023izd}. This measurement in full agreement with those from BES(III) was based on the $3\pi$ decay channel selected from $1.75\times 10^8$ tau pair events using the pseudo-mass endpoint method. The comparison shows nicely the complementarity of different experiments and colliders. 

\subsubsection{Tau lifetime measurements}
A measurement of the tau lepton lifetime $\tau_\tau$ needs sufficient boost before it decays. Therefore, contrary to $m_\tau$, it is unlikely that the STCF would provide a competitive measurement of $\tau_\tau$ using the traditional methods. At the $B$-factory, the mean decay length is around 250\,$\mu$m. Its measurement is thus possible using high resolution vertex and tracking detectors. At the $Z$-factory, the measurement was even easier as the mean decay length reaches to 2.2\,mm. This is why for many years, the measurements from the LEP experiments dominate. The best precision achieved by the Belle experiment in 2014 was $(290.17\pm 0.53\mathrm{(stat)}\pm 0.33\mathrm{(syst)})$\,fs~\cite{Belle:2013teo}.
Improved measurements are expected with Belle II at SuperKEKB thanks to the nano-beam scheme reducing the beam spot size at the interaction point and making it smaller than any of the previous collider experiments. 

The measurement of $\tau_\tau$ shows again the complementarity of different experiments and colliders. The precision measurements of $\tau_\tau$, $m_\tau$ and the tau decay branching fractions are needed for a number of important tests to be discussed later in the proceedings.  

\subsubsection{Tau decay branching fraction measurements}
Given its large mass, the tau lepton is the only lepton, among the three charged ones in the SM, heavy enough to have hadronic decay modes. The branching fractions (${\cal B}$) used to be measured individually by different experiments at different colliders, resulting in a longstanding puzzle of ``one-prong problem''~\cite{PhysRevD.30.1509,PhysRevD.31.1066}, namely, the ${\cal B}_1$ of the inclusive decay modes containing one-prong does not correspond to the sum of the exclusive decay modes.

A possible solution was indicated by the CELLO experiment~\cite{cello1990exclusive}.
The problem was finally resolved by the ALEPH's simultaneous measurement of all major decay modes first with the initial data sample of 1989 and 1990~\cite{ALEPH:1991jyg} and later with the full data sample of LEP-1~\cite{ALEPH:2005qgp}. The results are shown in Table~\ref{tab:br-aleph} as some of channels still have the best precision 20 years after its measurement.
\begin{table}[tbp]
\caption{ALEPH's results for the exclusive branching fractions ($\cal{B}$) 
         for modes without kaons~\cite{ALEPH:2005qgp}. The contribution of the kaon modes having a total value of $(2.87\pm 0.12)\%$ has been subtracted from the relevant decay modes~\cite{ALEPH:1999uux}.}
\begin{center}
\begin{tabular}{lrc}
\toprule
 mode & $\cal{B}\pm\sigma_{\hbox{stat}} \pm \sigma_{\hbox{syst}}$
  [\%] \\\colrule
 $e$ &    17.837 $\pm$  0.072 $\pm$  0.036 &\\
 $\mu$ &    17.319 $\pm$  0.070 $\pm$  0.032 &\\
 $\pi^-$ &    10.828 $\pm$  0.070 $\pm$  0.078 &\\
 $\pi^-\pi^0$ &    25.471 $\pm$  0.097 $\pm$  0.085 &\\
 $\pi^-2\pi^0$ &     9.239 $\pm$  0.086 $\pm$  0.090 &\\
 $\pi^-3\pi^0$ &      0.977 $\pm$  0.069 $\pm$  0.058 &\\
 $\pi^-4\pi^0$ &      0.112 $\pm$  0.037 $\pm$  0.035 &\\
 $\pi^-\pi^+\pi^-$ &     9.041 $\pm$  0.060 $\pm$  0.076 &\\
 $\pi^-\pi^+\pi^-\pi^0$ &     4.590 $\pm$  0.057 $\pm$  0.064 &\\
 $\pi^-\pi^+\pi^-2\pi^0$ &      0.392 $\pm$  0.030 $\pm$  0.035 &\\
 $\pi^-\pi^+\pi^-3\pi^0$ &      0.013 $\pm$  0.000 $\pm$  0.010 & Estimate\\
 $3\pi^-2\pi^+$ &      0.072 $\pm$  0.009 $\pm$  0.012 &\\
 $3\pi^-2\pi^+\pi^0$ &      0.014 $\pm$  0.007 $\pm$  0.006 &\\
$\pi^- \pi^0 \eta$ & 0.180 $\pm$ 0.040 $\pm$ 0.020 & ALEPH~\cite{ALEPH:1996kok}\\
$\pi^- 2\pi^0 \eta$ & 0.015 $\pm$ 0.004 $\pm$ 0.003 & CLEO~\cite{bergfeld1997first}\\
$\pi^- \pi^- \pi^+ \eta$ & 0.024 $\pm$ 0.003 $\pm$ 0.004 & CLEO~\cite{bergfeld1997first}\\
$a_1^- (\rightarrow \pi^- \gamma)$ & 0.040 $\pm$ 0.000 $\pm$ 0.020 & Estimate \\
$\pi^- \omega (\rightarrow \pi^0 \gamma, \pi^+ \pi^-)$ & 0.253 $\pm$ 0.005 $\pm$ 0.017 & ALEPH~\cite{ALEPH:1996kok}\\
$\pi^- \pi^0 \omega (\rightarrow \pi^0 \gamma, \pi^+ \pi^-)$ & 0.048 $\pm$ 0.006 $\pm$ 0.007 & ALEPH~\cite{ALEPH:1996kok} + CLEO~\cite{PhysRevLett.71.1791}\\
$\pi^- 2\pi^0 \omega (\rightarrow \pi^0 \gamma, \pi^+ \pi^-)$ & 0.002 $\pm$ 0.001 $\pm$ 0.001 & CLEO~\cite{bergfeld1997first}\\
$\pi^- \pi^- \pi^+\omega (\rightarrow \pi^0 \gamma, \pi^+ \pi^-)$ & 0.001 $\pm$ 0.001 $\pm$ 0.001 & CLEO~\cite{bergfeld1997first}\\
\botrule
\end{tabular}
\label{tab:br-aleph}
\end{center}
\end{table}

The ALEPH measurement has benefited from its good efficiency ($\sim 90$\%) with high acceptance for all the dominant channels. The hadronic background is easy to be separated from the signal based mainly on their different charge track multiplicity and is below 1\% after the selection. The particle identification among $e$, $\mu$ and hadrons (pion or kaon) is also excellent thanks to the good and redundant subdetectors. 

The STCF can and should play a role in improving the precision of the existing decay modes and detecting additional rare decays using the cutting-edge technology employed in the modern detector and the high statistics tau data samples. Near the tau pair production threshold, there is also the possibility of using monochromatic distributions for additional kinematic separation. The (light) quark background near threshold can be well controlled, measured and extrapolated. It remains to be checked whether the beam-related background can be reduced to a small and negligible level. 

\subsection{Lepton-flavor-universality tests}
By using the $m_\tau$, $\tau_\tau$ or the relevant $\cal{B}$, a number of LFU tests can be performed~\cite{Davier:2005xq}.
The relation between the three quantities in the SM is known~\cite{PhysRevLett.61.1815}
\begin{equation}
\frac{{\cal B}(\tau^-\to\nu_\tau\ell^-\overline{\nu}_\ell)}{\tau_\tau}=\frac{G_\tau G_\ell m^5_\tau}{192\pi^3}f\!\left(\frac{m^2_\ell}{m^2_\tau}\right)\delta^{\tau}_W \delta^\tau_\gamma
\label{eq:lfu}
\end{equation}
where $G_{\tau/\ell}=\frac{g^2_{\tau/\ell}}{4\sqrt{2}M^2_W}$, $f(x)=1-8x+8x^3-x^4-12x^2\ln x$ and $\delta^{\tau}_W=1+\frac{3}{5}\frac{m^2_\tau}{M^2_W}+\frac{9}{5}\frac{m^2_\ell}{M^2_W}$ and $\delta^\tau_\gamma=1+\frac{\alpha(m_\tau)}{2\pi}\left(\frac{25}{4}-\pi^2\right)$.
By replacing $\tau$ by $\mu$, the formula applies also to muon decay $\mu^-\to\nu_\mu e^-\overline{\nu}_e$. 
The heavy lepton mass dependence of the fifth power explains the huge difference in lifetime between leptons $\tau$ and $\mu$.

\begin{figure}[tb]
\centerline{\includegraphics[width=5.5cm]{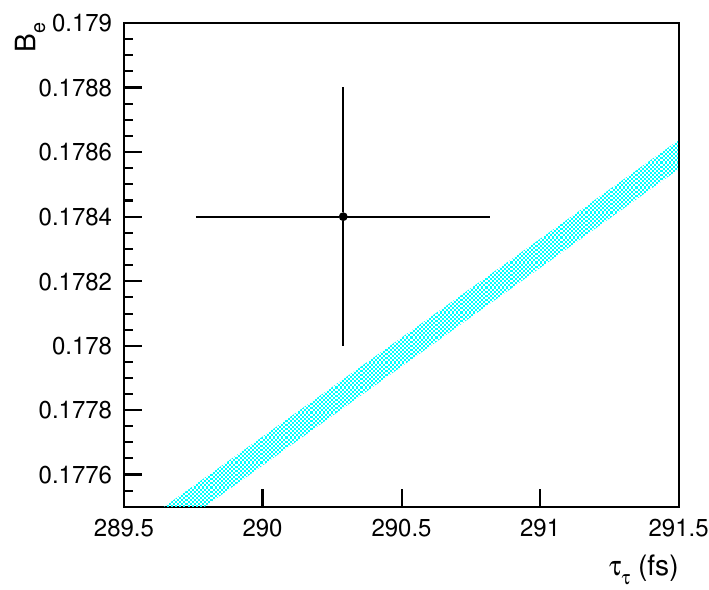}}
\caption{Measured $\tau_\tau$, ${\cal B}_e$ in comparison with the expected value using the measured $m_\tau$. These measured values correspond to averaged one from HFLAV~\cite{HeavyFlavorAveragingGroupHFLAV:2024ctg}. The size of the error band is due to the uncertainty of $m_\tau$.}
\label{fig:tau-lfu}
\end{figure}

One LFU test is to compare the measured ${\cal B}_e$ and $\tau_\tau$ with the expected one derived using the measured $m_\tau$ in Eq.~(\ref{eq:lfu}). 
The comparison is shown in Figure~\ref{fig:tau-lfu}.
The test is currently limited by the precision of ${\cal B}_e$ and $\tau_\tau$, whereas the precision of $m_\tau$, as reflected in the size of the shaded band, is adequate.
The STCF can improve the measurement of ${\cal B}_e$ and eventually also $m_\tau$.

Several other LFU tests can be performed by comparing either different decay modes of the $\tau$ decays or the same decay mode between the $\tau$ and $\mu$ decays. The LFU tests are important as they may unveil some clue on the puzzling family structure of matter.


\subsection{Rare and lepton-flavor-violating decays}

\begin{figure}[hbt]
\centerline{\includegraphics[width=5.5cm]{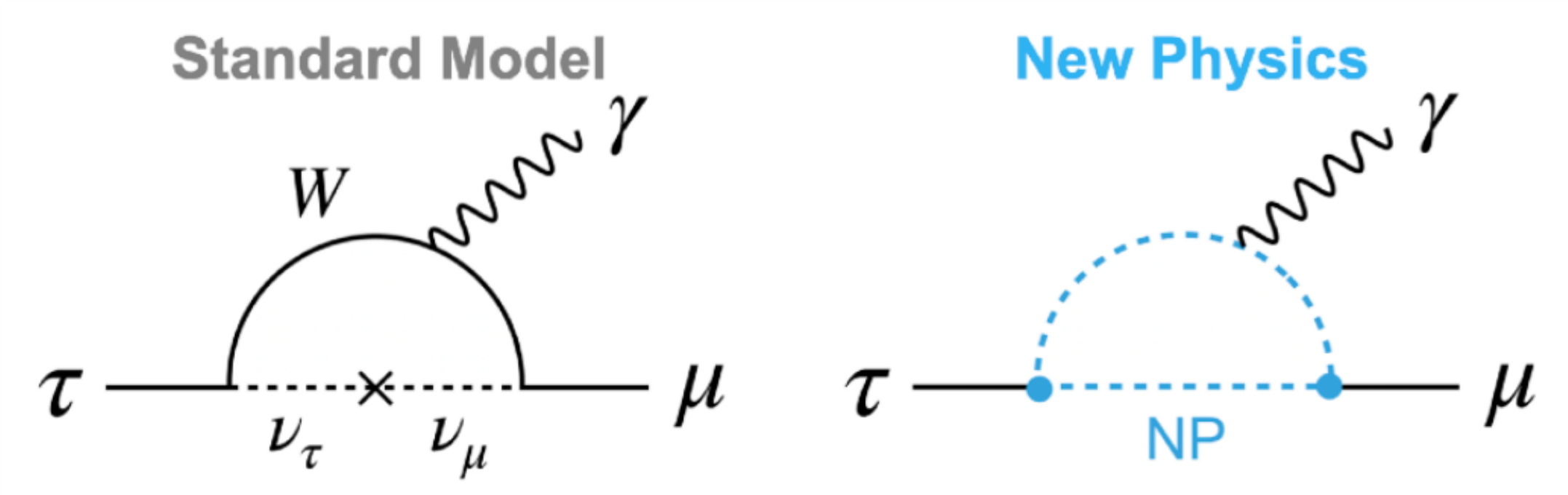}}
\caption{Graphs for tau lepton-flavor-violating decays from the SM and new physics contributions.}
\label{fig:graph-lfv}
\end{figure}

\begin{figure}[bt]
\centerline{\includegraphics[width=12cm]{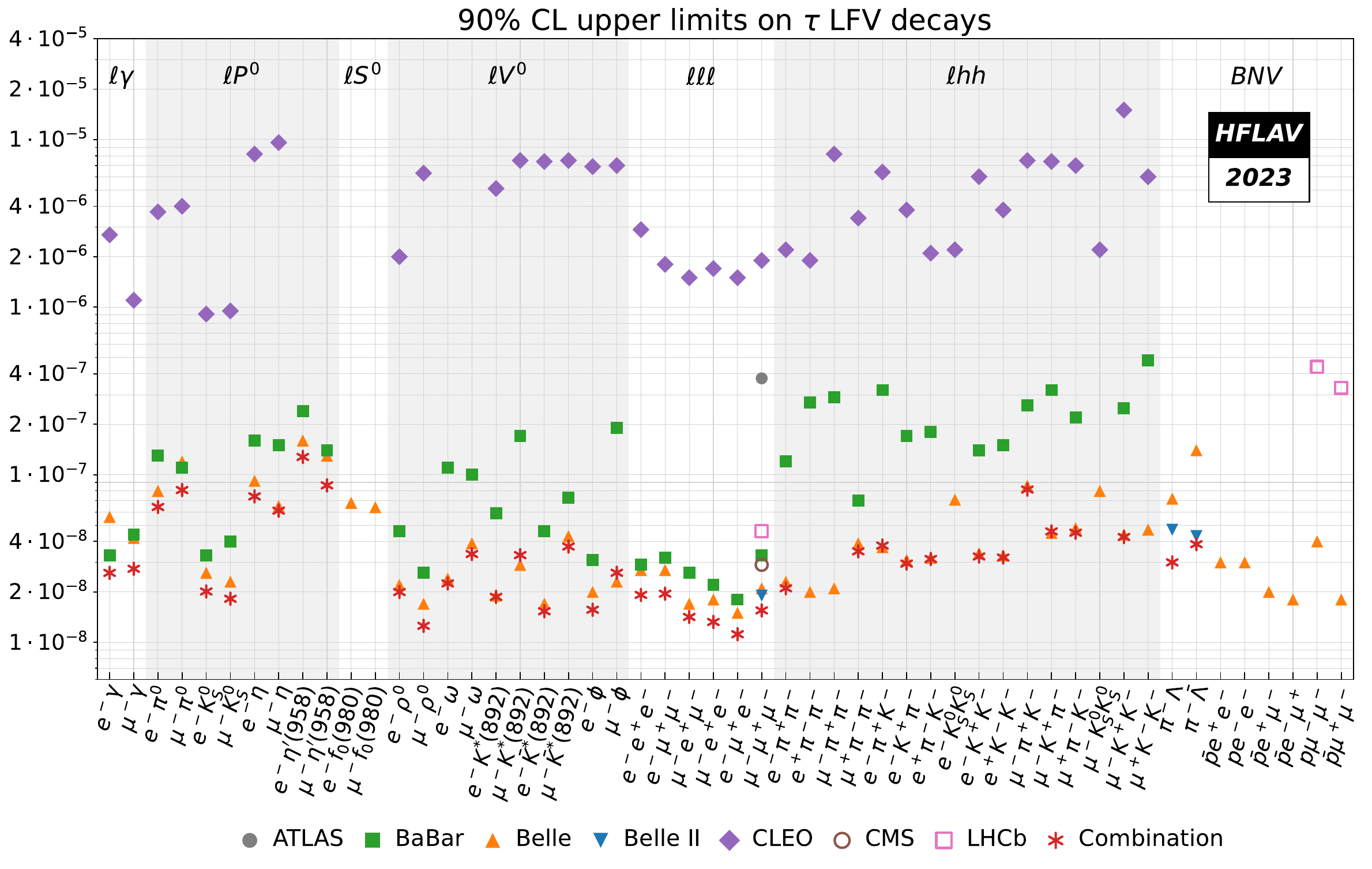}}
\caption{Tau lepton-flavor-violating branching fraction upper limits and combinations. Figure taken from Ref.~\cite{HeavyFlavorAveragingGroupHFLAV:2024ctg}.}
\label{fig:lfv-limits}
\end{figure}

There are many possible rare and LFV decay modes in tau decays. For instance, the decay mode $\tau\to \mu\gamma$ is possible in the SM due to the non-zero masses of the tau and muon neutrinos (Figure~\ref{fig:graph-lfv} left). The branching fraction is, however, proportional to $\left(\frac{\Delta m^2_\nu}{m^2_W}\right)^2<10^{-50}$ and thus negligible. New physics (Figure~\ref{fig:graph-lfv} right) may also contribute. Any observation for such a LFV decay mode would be a direct signal for new physics. This explains why the LFV searches have been a major research topic for all relevant experiments and there are good prospects for the tau factories (Figure~\ref{fig:lfv-limits}).   

There has been a dedicate STCF sensitivity study~\cite{Xiang:2023mkc} using a fast simulation in the decay mode $\tau\to \mu\gamma$. The upper limit at 90\% CL is estimated to $2.8\times 10^{-8}$ and $8.8\times 10^{-9}$ for an integrated luminosity of 1\,ab$^{-1}$ and 10\,ab$^{-1}$, respectively. The sensitivity is thus competitive to the current combined limit and future sensitivity from Belle\,II.

There are many competing experiments plus those for non tau processes, e.g.\ Mu2e@FermiLab, Mu3e@PSI, MEGII@PSI, COMET@J-PARC. STCF should not be too much delayed in order not to miss a chance.

\subsection{Tau electric dipole and magnetic moment anomaly}
\subsubsection{Tau electric dipole moment}\label{sec:dtau}
The tau electric moment ($d_\tau$) characterizes the time-reversal (T) or charge-parity (CP) violation properties at $\gamma\tau\tau$ vertex.
The general Lagrangian describing the CP/T violation in tau pair production related to the electric dipole moment (and weak dipole moment, $d_\tau^W$) is
\begin{equation}
{\cal L}_\mathrm{CP/T} = -\frac{1}{2}i\overline{\tau}\sigma^{\mu\nu}\gamma_5\tau\left[d_\tau(q^2)F_{\mu\nu}+d^W_\tau(q^2)Z_{\mu\nu}\right]
\,,
\label{diseqn}
\end{equation}
where $F_{\mu\nu}$ and $Z_{\mu\nu}$ are the electromagnetic and weak field tensors. At the STCF, the momentum transfer $q$ is around 4\,GeV/c, $d_\tau^W$ in the second term is a factor $\frac{4m^2_\tau}{M^2_Z}\simeq 2\times 10^{-3}$ than $d_\tau$ and can be safely neglected. 

In the SM, the CP violation arises due to an irreducible complex phase in the CKM matrix~\cite{Kobayashi:1973fv}, which predicts an extremely small value, $d_\tau\approx 10^{-37}\,e$cm~\cite{Booth:1993af,Mahanta:1996er}, many orders of magnitude below any experimental sensitivity.
A large class of new physics models predicts enhanced contributions in $d_\tau$ at observable levels of $10^{-19}\,e$cm~\cite{Huang:1996jr}.

Independent measurements of the electric dipole moments of $e$, $\mu$ and $\tau$ are necessary to determine the flavor dependence of CP violating phases in the possible mixing between three generation of the charged lepton sector.
In general, the strength of CP violation may be different for different flavors. 
The current best experimental sensitivity for $d_\tau$ is around $10^{-17}\,e$cm~\cite{Belle:2021ybo}. 
This is to be compared with those for $d_e$ and $d_\mu$ of $10^{-30}\,e$cm~\cite{Roussy:2022cmp} and $10^{-19}\,e$cm~\cite{Muong-2:2008ebm}, respectively.

A recent sensitivity study for the STCF has been performed based on the two-pion decay mode selected from a simulated MC sample of around $5.5\times 10^6$ tau pair events at $\sqrt{s}=4.68$\,GeV~\cite{Sun_2025}. The sensitivity obtained after extrapolating to a larger sample corresponding to 10 years operation is $d_\tau<3.89\times 10^{-18}\,e$cm at 68\% CL~\footnote{It would be good to change the sensitivity number at 95\% CL as those quoted above by the other determinations.}.

\subsubsection{Tau magnetic moment anomaly}

The tau lepton magnetic moment $\mu_\tau$, as $\mu_e$ and $\mu_\mu$ for the electron and muon, is an intrinsic property of an elementary charged lepton $\ell$.
It is proportional to its spin vector $\vec{S}_\ell$ as
\begin{equation}
\vec{\mu}_\ell= g_\ell \frac{q_\ell}{2m_\ell}\vec{S}_\ell
\end{equation}
where the $g_\ell$ factor predicted by Dirac about a century ago has a value of 2~\cite{dirac1928quantum}, $q_\ell$ and $m_\ell$ are the electric charge and mass of the lepton. Quantum corrections of the SM from all three sectors modify the value of $g_\ell$ by
2 per mil. The magnetic moment anomaly $a_\ell$ was introduced to quantify the deviation from 2:
\begin{equation}
a_\ell =\frac{g_\ell-2}{2}\,.
\end{equation}
The $a_e$ and $a_\mu$ are the two most precisely measured and predicted observables in particle physics.
Any deviation between the measurements and the predictions probes contributions beyond the SM.
Since the BSM contribution is in general proportional to $(m_\ell/\Lambda)^2$ with $\Lambda$ being the new physics scale, the tau lepton should be more sensitive to new physics contributions.  
Unfortunately its short lifetime precludes the same measurement methods used for the other two leptons. 
This explains why the current experimental sensitivity for $a_\tau$ is limited to $10^{-3}$~\cite{CMS:2024qjo}, well below those of $a_e$ and $a_\mu$ of $10^{-13}$~\cite{Fan:2022eto} and $10^{-10}$~\cite{Muong-2:2025xyk}, respectively.

In order to be sensitive to BSM effects, an improved experimental sensitivity for $a_\tau$ down to $10^{-6}$ is needed. This is very challenging and a longitudinally polarized beam can help to improve the sensitivity.

\subsection{CP violation in tau decays}
CP violation observed in the quark sector alone cannot explain the baryon asymmetry of the universe. The hadronic decays of the tau lepton provide a good opportunity to study and search for new CP violation in the lepton sector, in addition to the electric dipole moment related CP violation from the tau pair production, discussed in Section~\ref{sec:dtau}. Three different observables are studied in literature~\cite{Kilian:1994ub,TSAI1996272,Kiers:2008mv}: the integrated or full rate asymmetry, the differential angular distribution asymmetry and the triple-product asymmetry.

CP asymmetry in $\tau\to \pi K^0_S\nu_\tau$ with $K^0_S\to \pi^+\pi^-$
\begin{equation}
{\cal A}_\tau = \frac{\Gamma(\tau^+\to\pi^+ K^0_S\overline{\nu}_\tau)-\Gamma(\tau^-\to\pi^- K^0_S\nu_\tau)}{\Gamma(\tau^+\to\pi^+ K^0_S\overline{\nu}_\tau)+\Gamma(\tau^-\to\pi^- K^0_S\nu_\tau)}
\end{equation}
is expected\footnote{The $\pi$ final state can also be $K$, $\pi\pi$ etc.} due to the $K-\overline{K}$ mixing and predicted to be $+0.33(0.01)\%$~\cite{Bigi:2005ts,Grossman:2011zk}.
This process can be served as a calibration for searching for additional CP violation in tau decays. 
The measurement of BABAR of $-0.36(0.23)(0.11)\%$, with the first and second uncertainties being statistical and systematic, is in tension with the prediction by $2.8\sigma$. Independent measurements are needed to cross check the result of BABAR. 

A feasibility study at the STCF has been performed~\cite{Sang_2021} by comparing the decay rate difference between $\tau^+\to K_S\pi^+\overline{\nu}_\tau$
and $\tau^-\to K_S\pi^-\nu_\tau$ using a simulated MC sample corresponding to 1\,ab$^{-1}$ at $\sqrt{s}=4.26$\,GeV. The statistical sensitivity for CP violation is determined around $9.7\times 10^{-4}$. The sensitivity is improved to $3.1\times 10^{-4}$ with an integrated luminosity of 10\,ab$^{-1}$. See also Ref.~\cite{Cheng:2025kpp}.

\section{Hadronic decays, spectral functions and its applications}

Hadronic tau decays represent a clean laboratory for the precise study of quantum chromodynamics (QCD)~\cite{Davier:2005xq} and for constraining new physics~\cite{Cirigliano:2021yto}.
A large number of studies can be performed through the spectral functions, which embody both the rich hadronic structure seen at low energy and the quark behavior relevant in the higher energy regime. The spectral functions, playing an important role on the understanding of hadronic dynamics in the intermediate energy range, represent the basic input for evaluating low-energy contributions from hadronic vacuum polarization, for a precise extraction of the strong coupling at a relatively low energy scale, for determining the CKM element $|V_{us}|$ and the $s$-quark mass.

\subsection{Measurements of the spectral functions}

The spectral functions for a non-strange ($|\Delta S|=0$) or strange ($|\Delta S|=1$) vector (axial-vector) hadronic tau decay $V^-\nu_\tau$ ($A^-\nu_\tau$) is defined as
\begin{eqnarray}
    v_1(s)/a_1(s)&=& \frac{m^2_\tau}{6|V_\mathrm{CKM}|^2S_\mathrm{EW}}\frac{{\cal B}_{\tau^-\to V^-/A^-\nu_\tau}}{{\cal B}_{\tau^-\to e^-\overline{\nu}_e\nu_\tau}}\frac{dN_{V/A}}{N_{V/A}ds}\left[\left(1-\frac{s}{m^2_\tau}\right)^{\!\!2}\left(1+\frac{2s}{m^2_\tau}\right)\right]^{-1}\\
    a_0(s)&=& \frac{m^2_\tau}{6|V_\mathrm{CKM}|^2S_\mathrm{EW}}\frac{{\cal B}_{\tau^-\to \pi^-(K^-)\nu_\tau}}{{\cal B}_{\tau^-\to e^-\overline{\nu}_e\nu_\tau}}\frac{dN_{A}}{N_{A}ds}\left(1-\frac{s}{m^2_\tau}\right)^{\!\!-2}\,,
\end{eqnarray}
where $S_\mathrm{EW}$ accounts for short-distance radiative corrections~\cite{PhysRevLett.61.1815} and $\frac{dN_{V/A}}{N_{V/A}ds}$ is the normalized invariant mass distribution of the hadronic final state and for the pion (kaon) pole, it is a delta function $\delta(s-m^2_{\pi, K})$. 

The spectral function for the dominant $\pi\pi^0$ mode has been measured by ALEPH~\cite{ALEPH:2005qgp}, Belle~\cite{Belle:2008xpe}, CLEO~\cite{CLEO:1999dln} and OPAL~\cite{OPAL:1998rrm} under very different experimental conditions. 
ALEPH has the best normalization (branching fraction) measurement while Belle has the best shape measurement.
ALEPH has also measured a few other decay modes: $3\pi$, $\pi 2\pi^0$, $\pi 3\pi^0$, $\pi 4\pi^0$, $3\pi\pi^0$,  $3\pi 2\pi^0$ and $5\pi$. These are the dominant vector and axial-vector decay modes.

\subsection{Muon $g-2$ predictions}
The use of tau spectral functions for the leading-order (LO) hadronic vacuum polarization (HVP) evaluation was originally proposed in Ref.~\cite{Alemany:1997tn}. In the limit of isospin invariance, the vector current is conserved, so that the spectral functions of the vector tau decay modes in a given isospin state for the hadronic system are related to the $e^+e^-$ annihilation cross section of the corresponding isovector final state:
\begin{eqnarray}
    \sigma^{I=1}_{e^+e^-\to \pi^+\pi^-}&=& \frac{4\pi\alpha^2}{s}v_{1, \pi^-\pi^0\nu_\tau}\\
    \sigma^{I=1}_{e^+e^-\to 2\pi^+2\pi^-}&=& \frac{4\pi\alpha^2}{s}2v_{1, \pi^-3\pi^0\nu_\tau}\\
    \sigma^{I=1}_{e^+e^-\to \pi^+\pi^-2\pi^0}&=& \frac{4\pi\alpha^2}{s}\left[v_{1, 2\pi^-\pi^+\pi^0\nu_\tau}-v_{1, \pi^-3\pi^0\nu_\tau}\right]\,.
\end{eqnarray}
Using $e^+e^-\to$ hadrons cross sections in terms of ratio $R(s)$ of the hadronic cross sections over the point like $e^+e^-\to \mu^+\mu^-$ cross section in the dispersion relation
$a_\mu^\text{HVP LO}=\frac{\alpha^2}{3\pi}\int^\infty_{4m^2_\pi}ds\frac{K(s)}{s}R(s)$
with $\alpha$ being the QED coupling constant and $K(s)$ a QED kernel that has an energy dependence close to $1/s$, the LO HVP prediction can be obtained.
The same hadronic cross sections are also needed in obtaining the hadronic contribution to the running $\alpha(s)$ except that the QED kernel $K(s)$ has a different energy dependence. 

The tau data provide thus an alternative prediction for the LO HVP component of $a_\mu$ once the various isospin breaking corrections are taken into account~\cite{Davier:2010fmf,Davier:2023fpl}. The prediction is limited by the precision of the LO HVP contribution.
This is particularly important given that there are currently large discrepancies among the cross section measurements from different experiments for the dominant $e^+e^-\to \pi^+\pi^-$ channel which contributes about 73\% to the LO HVP. 

Nevertheless the precision of the tau based predictions is still limited. It is important that the STCF can help to improve the spectral function measurements as well as the relevant measurements for the isospin breaking corrections, e.g. the line shape differences between the charged and neutral rho mesons~\cite{Davier:2025jiq}.

\subsection{Precise determination of the strong coupling constant $\alpha_s$}

\begin{figure}[bt]
\centering
\includegraphics[width=6cm]{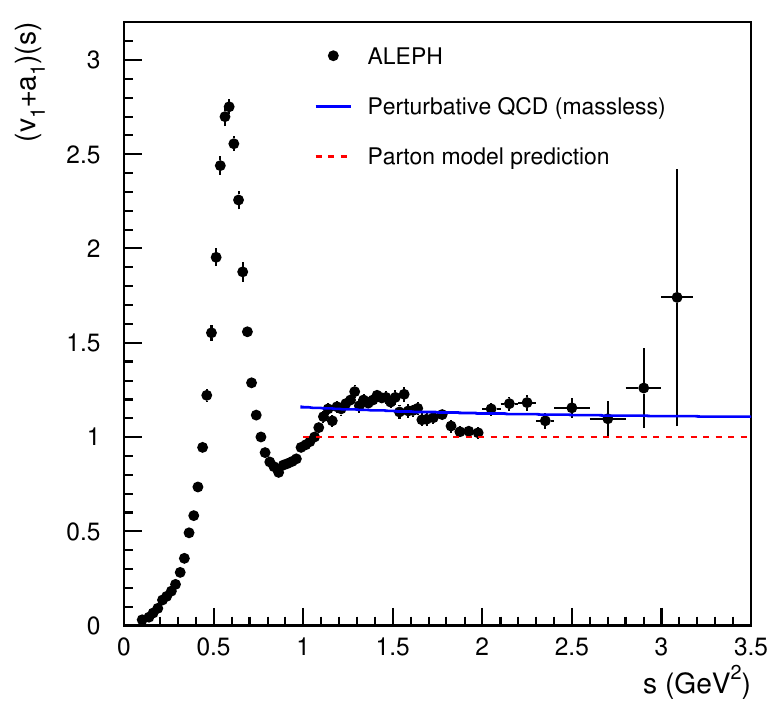}
\includegraphics[width=6cm]{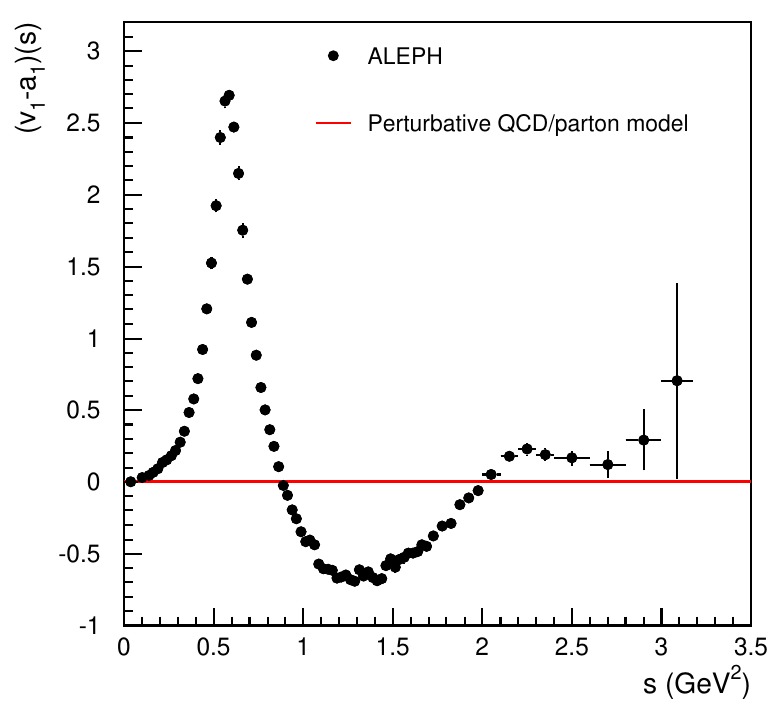}
\caption{Measured $V+A$ and $V-A$ spectral functions from ALEPH. Figures taken from Ref.~\cite{Davier:2013sfa}.}
\label{fig:sfs}
\end{figure}

Using the inclusive $v_1$ and $a_1$ spectral functions measured by ALEPH, the $V+A$ and $V-A$ spectral functions are shown in Figure~\ref{fig:sfs}.
It is interesting to note that despite the limited precision at the high energy tail, the $V+A$ spectral function reaches the asymptotic limit in agreement with the perturbative QCD prediction. This provides the basis for a precise determination of $\alpha_s$ at an energy scale down to $m_\tau$, whereas the $V-A$ spectral function receives purely non-perturbative contribution and shows strong oscillation around the parton model prediction as the low energy part of the $V+A$ spectrum.

The extracted value is $\alpha_s(m^2_\tau)=0.332(5)(11)$ with the first and second uncertainties being experimental and theoretical~\cite{Davier:2013sfa}. 
The value after evolution to $M_Z$ is $\alpha_s(M^2_Z)=0.1199(6)(12)(5)$ with the third uncertainty due to the evolution. This result is in excellent agreement with the determination from the $Z$ width of $\alpha_s(M^2_Z)=0.1186(27)$ based on the global fit to all electroweak data~\cite{LEP:2004xhf}. The consistency between the two results provides a powerful test of the evolution of $\alpha_s(s)$ as predicted by the non-abelian nature of the QCD gauge theory.

\subsection{Strange spectral functions and its applications}

\begin{figure}[hbt]
\centering
\includegraphics[width=12cm]{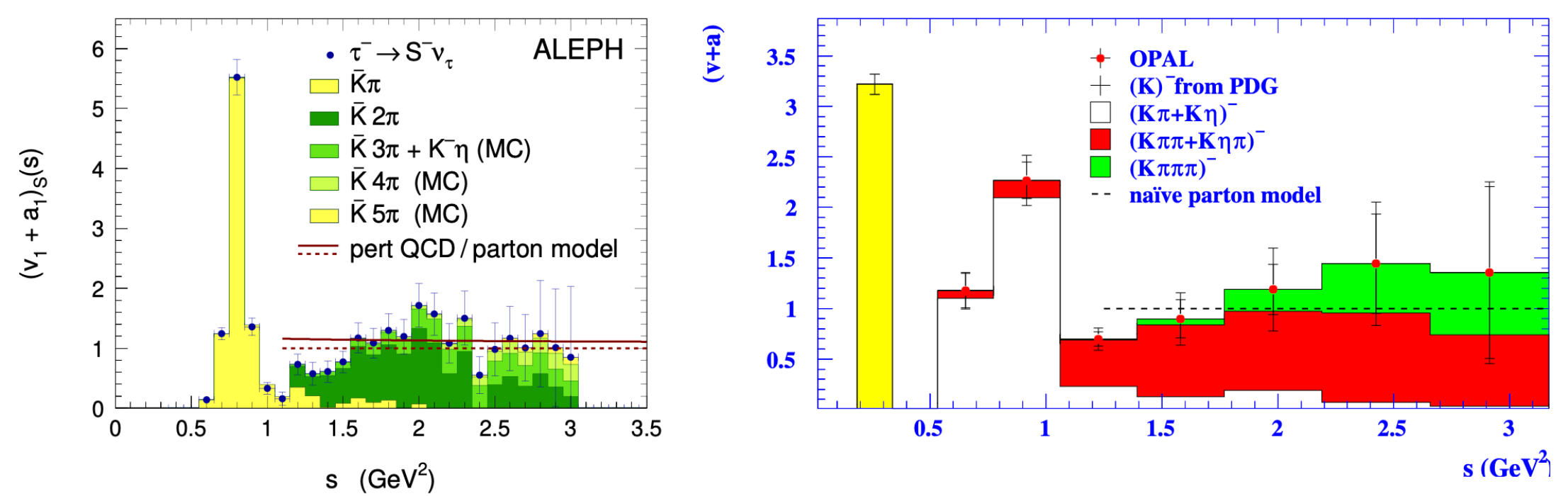}
\caption{Inclusive $V+A$ spectral function from tau decays into $S=\pm 1$ final states from ALEPH (left) and OPAL (right). The kaon pole is not included in the ALEPH plot. Figures taken from Ref.~\cite{Davier:2005xq}.}
\label{fig:s-sfs}
\end{figure}

The measured strange spectral functions are presented in Figure~\ref{fig:s-sfs}. It should be noted that the OPAL measurement was normalized with branching fraction values from PDG, which were dominated by the ALEPH measurements. Belle has also measured the $K_S\pi$ spectral function in 2007~\cite{Belle:2007goc}. 
The spectral functions can be used to extract $|V_{us}|$ and the strange quark mass $m_s$~\cite{Davier:2005xq}. This is important in particular in view of the current discrepancy between the various $|V_{us}|$ determinations and the unitarity result~\cite{HeavyFlavorAveragingGroupHFLAV:2024ctg}.

In comparison with the non-strange spectral functions, the precision is very limited. The STCF should be able to play an important role here if the situation will not be not improved.

\section{Other topics}
There are a few other important topics which are not discussed in the presentation due to the time constraint. These include dark sector searches for dark matter or axion-like candidates~\cite{Jiang:2025nie}; improved measurements of tau decay parameters -- Michel parameters -- to study if the Lorentz structure of charged weak currents is SM-like; the investigation of quantum entanglements; the prospects for ditauonium discovery, which would provide the most precise determination of $m_\tau$~\cite{dEnterria:2023yao}; and searches for new physics using effective filed theory approaches. 

\section{Summary}

As far as the tau physics is concerned, the STCF is unique for some of the observations and measurements. Most of the observations and measurements can, however, be realized at all colliders. In this sense, we should not wait too long before constructing the STCF, as the other existing experiments will not wait for the STCF. It is also important to have different experiments so that any new observation can be verified. 
\begin{figure}[hbt]
\centering
\includegraphics[width=12cm]{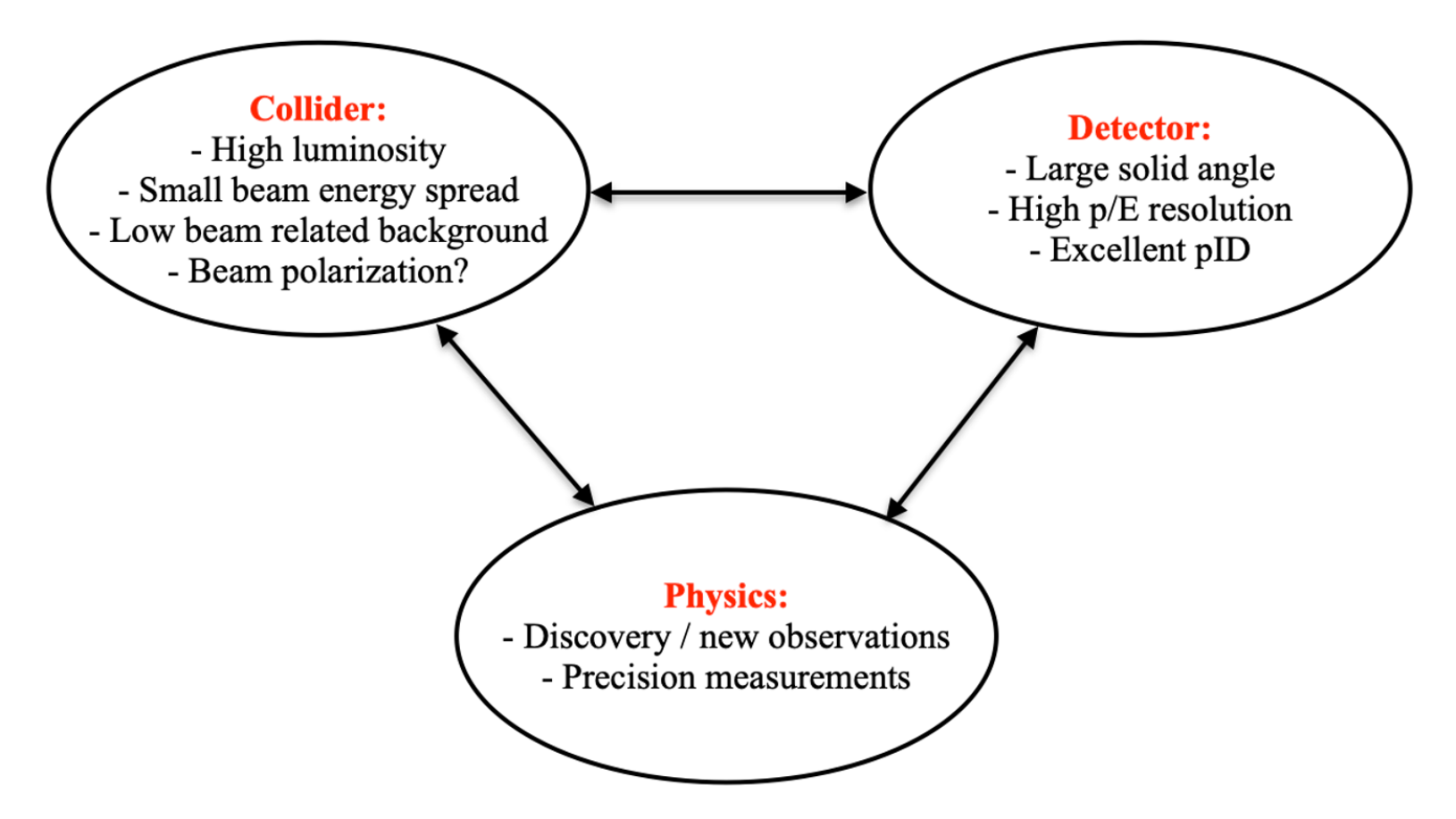}
\caption{Strong interplay between the collider, detector and physics results.}
\label{fig:interplay}
\end{figure}

There is a strong interplay between the collider to be built, the detector to be constructed and the physics results to be achieved as indicated in Figure~\ref{fig:interplay}. It would be good to have the best accelerator and detector possible from the beginning and make them upgradable. This would maximize the outcome of the physics results. It is also desirable to perform more feasibility studies for accelerator and detector optimization.

\bibliographystyle{unsrt}
\bibliography{main}

\end{document}